\documentclass[prb,aps,twocolumn,floatfix]{revtex4}

\usepackage{amsmath,amsfonts,amssymb}
\usepackage{wrapfig}
\usepackage{graphicx}
\usepackage{bbm}
\usepackage{graphics}

\def \beq {\begin{equation}}
\def \eeq {\end{equation}}
\def \ba {\begin{eqnarray}}
\def \ea {\end{eqnarray}}

\begin{document}

\title{Observation of optomechanical buckling phase transitions} 
\author{H. Xu,$^{1}$, U. Kemiktarak$^{1,2,}$, J. Fan,$^{2}$, S. Ragole$^{1,3}$, J. Lawall,$^{2\ast}$, J. M. Taylor$^{1,2,3\ast}$}
\affiliation{
\normalsize{$^{1}$Joint Quantum Institute, University of Maryland, College Park, MD  20742, USA}\\
\normalsize{$^{2}$National Institute of Standards and Technology, Gaithersburg, MD 20899, USA}\\
\normalsize{$^{3}$Joint Center for Quantum Information and Computer Science,} \\
\normalsize{University of Maryland, College Park, MD 20742, USA}\\
}

\begin{abstract}
Correlated phases of matter provide long-term stability for systems as diverse as solids, magnets, and potential exotic quantum materials. Mechanical systems, such as relays and buckling transition spring switches can yield similar stability by exploiting non-equilibrium phase transitions.  Curiously, in the optical domain, observations of such phase transitions remain elusive\cite{dorsel_optical_1983,meystre_theory_1985}. However, efforts to integrate optical and mechanical systems -- optomechanics\cite{milburn_introduction_2011,meystre_short_2013,aspelmeyer_cavity_2014}  -- suggest that a hybrid approach combining the quantum control of optical systems with the engineerability of mechanical systems may provide a new avenue for such explorations\cite{metcalfe_applications_2014,abramovici_ligo:_1992,li_harnessing_2008,ludwig_enhanced_2012,winger_chip-scale_2011}.  Here we report the first observation of the buckling of an optomechanical
system, in which transitions between stable mechanical states corresponding to both first- and second-order phase transitions are driven by varying laser power and detuning. Our results enable new applications in photonics and, given rapid progress in pushing optomechanical systems into the quantum regime, the potential for explorations of quantum phase transitions.
\end{abstract}

\maketitle

Optomechanical systems provide a unique connection between light and mechanical motion\cite{milburn_introduction_2011,meystre_short_2013,aspelmeyer_cavity_2014} due
to both their conceptual simplicity -- radiation pressure force
induces motion in a compliant optical element -- and their practical
applications in photonics and sensing\cite{metcalfe_applications_2014,abramovici_ligo:_1992,li_harnessing_2008,ludwig_enhanced_2012,winger_chip-scale_2011,krause_high-resolution_2012}.
A canonical example is the modification of the mechanical
spring constant via dynamical effects from the optical modes coupled
to the mechanical system.  First demonstrated experimentally in~2004\cite{sheard_observation_2004},
the so-called ``optical spring'' effect has been employed in the contexts of gravity-wave
detection\cite{VirgoOptSpring2006}, optical trapping of a mirror\cite{corbitt_all-optical_2007}, raising the mechanical quality factor (``optical dilution'')
of a mechanical oscillator\cite{corbitt_optical_2007}, and optical cooling\cite{CoolingSpring2008}.
At the same time, more complex mechanical elements provide new opportunities.
For example, nanomechanical devices can be used as memory cells\cite{mahboob_bit_2008,badzey_controllable_2004,bagheri_dynamic_2011}
or as logic gates. A crucial ingredient for these applications is
to develop a robust element that, when driven electronically or optically,
can be set to one of two stable static states. 
The first experimental demonstration of bistability induced by radiation pressure
was performed by Dorsel and collaborators in 1983\cite{dorsel_optical_1983}, in which the 
length, and thus the resonant frequency, of a Fabry-Perot cavity was modified by the 
circulating optical power. Shortly thereafter an analogous experiment was performed
in the microwave domain\cite{gozzini_light-pressure_1985}.  This work was followed 
by numerous proposals of applications,
including the realization of controllable buckled optomechanical systems\cite{meystre_theory_1985}.
Somewhat surprisingly, the applications were not pursued, and in fact to our knowledge the
only experiments involving optomechanical bistability reported in the meantime have either involved
additional electrostatic feedback\cite{Mueller2008} or a photothermal mechanism\cite{metzger_self-induced_2008,Hao_photothermal_bistable} rather than radiation pressure.
Instead, dynamical effects, necessary for laser cooling and exploration of narrowband
behavior, have been the focus of researchers in nanoscale optomechanics in recent years.
At the same time, static properties beyond bistability provide a new
window\cite{meystre_theory_1985,buchmann_macroscopic_2012} into
both our understanding of phase transitions and new application spaces
for optomechanics in sensing and optical information processing.
While some experiments with optically
driven, pre-buckled devices have yielded successes\cite{bagheri_dynamic_2011},
the mechanical potential was not optically modified in those systems.

Here we report the observation of radiation-pressure induced buckling phase transitions in an
optomechanical system for the first time since it was predicted
several decades ago\cite{meystre_theory_1985}.  
Our approach relies on a symmetrical optical cavity with a dielectric membrane
in the middle\cite{thompson_strong_2008,jayich_dispersive_2008,sankey_strong_2010}, where the behavior of the mechanical system can be fabricated and characterized separately from the optics
necessary to realize an optical cavity of high finesse.  Using this platform, we demonstrate an
optomechanical system that realizes controlled first-~and second-order
buckling transitions, where the overall displacement of the membrane is the order parameter. These transitions can be understood as arising when the static
optomechanical potential changes smoothly from a single well to a multiwell
potential as the optical driving power is increased.  Unlike the situation in the pioneering experiment
of Dorsel, in which the bistability was associated with a manifestly asymmetric optomechanical potential\cite{meystre_theory_1985},
our realization results in a spontaneously broken symmetry as the optical
drive passes through the transition point.
We derive the phase diagram for the buckling transitions, and show good quantitative agreement between crucial points predicted in our phase diagram and the experiment.

\begin{figure}
\includegraphics[width=3in]{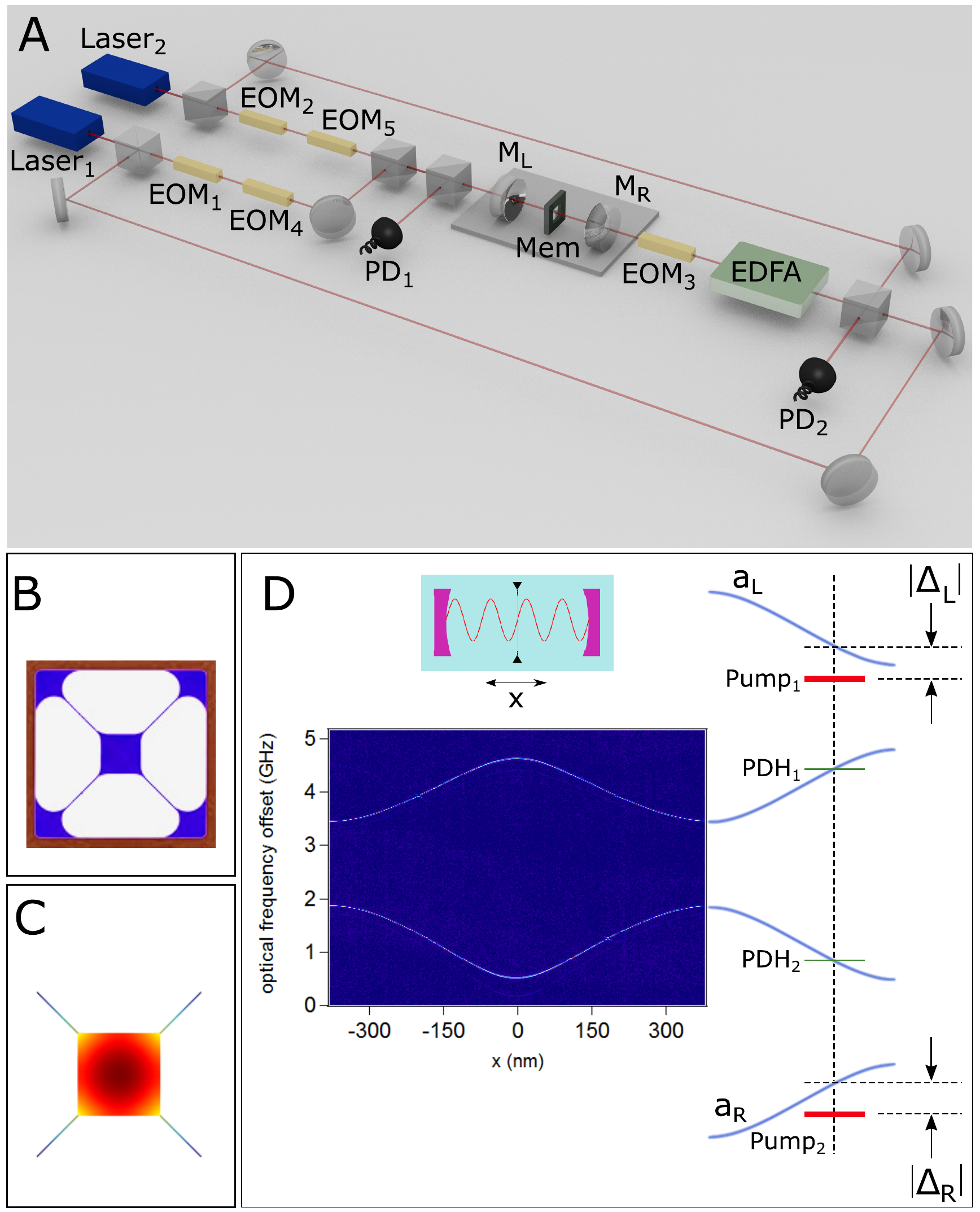}
\caption{\textbf{Experimental setup}\newline
\textbf{a,} Curved mirrors $\mathrm{M_L}$ and $\mathrm{M_R}$
form a symmetric Fabry-Perot cavity. The tethered membrane that comprises our
optomechanical resonator is placed in the center. 
The two pump and two probe fields used in the experiment are generated
and locked to the cavity by means of independent tunable lasers 
$\mbox{Laser}_{1}$ and $\mbox{Laser}_{2}$, electro-optic phase modulators
$\mbox{EOM}_{1}$--$\mbox{EOM}_{5}$, and an erbium-doped fiber amplifier
EDFA  (details are given in the Appendix).
Light reflected from the cavity is captured by photodiode
PD$_1$ in order to lock the lasers to the cavity, and the beat signal
between $\mbox{Laser}_{1}$ and $\mbox{Laser}_{2}$ is captured on PD$_2$ 
to probe the membrane position.
\textbf{b,} Microscope image of tethered SiN membrane.  The central square is 200~$\mu$m on a side.
\textbf{c,} Fundamental mechanical mode of membrane determined from finite element analysis; the frequency
is 80.3~kHz.
\textbf{d,} Optical transmission spectrum for membrane position $x$ near the center of
the cavity. Probe lasers PDH$_1$ and PDH$_2$ are locked to adjacent cavity
modes whose resonance frequencies have opposite dependences on membrane displacement. 
Pump laser Pump$_1$ is rigidly offset to probe PDH$_1$ and detuned to the red of an adjacent
cavity mode by  $\Delta$, and similarly for Pump$_2$, as shown.  When the membrane is displaced 
to the right, Pump$_1$ is brought closer to resonance and Pump$_2$ is driven further from
resonance.}
\end{figure}

A schematic of our apparatus is shown in Fig.~1a. Two dielectric mirrors, identified as $\mathrm{M_L}$ and $\mathrm{M_R}$, form an optical cavity
with a length of $\sim50$~mm. At the center of the cavity is the optomechanical element, a tensioned silicon nitride membrane normal to the cavity mode with long thin tethers connecting the membrane to its frame. An image made with an optical microscope is
shown in Fig.~1b, and a finite-element simulation of the fundamental 
mechanical mode is shown in Fig.~1c. It has a fundamental mechanical frequency of $\omega_m=2\pi\times 80.3$~kHz, designed to allow for substantial optical spring effects at low laser power. 
The experiment is performed at room temperature and pressure.  

Our approach relies on driving two different optical modes, denoted $a_L$ and $a_R$,
which have optomechanical couplings of opposite signs.   For small displacements
of the membrane to the left, mode $a_L$ is shifted up in frequency while $a_R$ is 
shifted down, and vice-versa.  A two-dimensional experimental plot of cavity transmission versus
membrane position and optical frequency is shown in Fig.~1d; the 
spectrum is periodic with frequency, so for any membrane positions~$x$
except those corresponding to extrema in the spectrum, there is a
large set of pairs of modes with opposite optomechanical couplings $g_{L,R}$ (slope
of curves in Fig.~1d).
The linewidth (FWHM) $\kappa/(2\pi)$ of the cavity modes is approximately 1.8 MHz.
We move the membrane about 55 nm away from an anti-crossing, where
the optomechanical couplings have amplitudes of $g_{L,R}=\pm2\pi \times 2.1$~kHz/pm,
and use two such pairs, as illustrated to the right of Fig.~1d.
Conceptually, there are four laser fields involved in our experiment,
as we now describe.

Two independent probe lasers are locked to a pair of modes with
opposite optomechanical couplings by means of the Pound-Drever-Hall
(PDH) method, and denoted PDH$_1$ and PDH$_2$ in Fig.~1d. 
The probe fields are actually first-order sidebands generated by electro-optic 
phase modulators $\mbox{EOM}_{1}$ and $\mbox{EOM}_{2}$, as shown in Fig.~1a, on independent lasers 
denoted ``Laser$_1$'' and ``Laser$_2$.''   (A detailed description of
how the laser fields are generated can be found in the Appendix).  
Due to the opposite signs of the optomechanical couplings
of the modes to which the probe fields are locked, the frequency difference
between PDH$_1$ and PDH$_2$ is proportional to the membrane displacement,
with twice the response of either mode alone.  
We access this frequency difference by counting the beat signal between
the two lasers as detected with photodetector PD2.  At the 2~kHz data acquisition rate employed
in this work, the position measurement resolution is below 1~pm.

We supplement the probe fields with additional fields in order to create
a tailorable multiwell optomechanical potential.
Two strong pump fields, denoted Pump$_1$ and Pump$_2$ in  Fig.~1d, are generated by combining light 
from the lasers, amplifying it, and passing it through phase modulator $\mbox{EOM}_{3}$.
Pump$_1$ has power $P_L$  and (angular) frequency $\nu_L$ and is frequency-offset from probe field PDH$_1$ by $\sim2$~GHz, the sum of the
drive frequencies for $\mbox{EOM}_{1}$ and $\mbox{EOM}_{3}$, 
such that it drives mode $a_L$ with (at low pump power) red detuning $\Delta_L$.
Pump$_2$ has power $P_R$  and angular frequency $\nu_R$ and is similarly offset from probe PDH$_2$ so as to drive mode 
$a_R$ with detuning $\Delta_R$.  Crucially, the optomechanical
coupling of each pump field is of the opposite sign of the probe field to 
which it is frequency-offset.  While a rich variety of phenomena is accessible by
taking independent values of $P_L$, $\Delta_L$, $P_R$ and $\Delta_R$, the experiments
described here employ the symmetric situation $P_L=P_R$ and $\Delta_L=\Delta_R$.

In this configuration, there is only one stable steady state for low to intermediate pump power levels. At higher powers, however, additional steady states appear. For the symmetric case we have constructed, 
the solutions to the dynamical equations describing these steady states have a particularly simple form. 
The optical fields $a_L$ (left-moving) and $a_R$ (right-moving) take the form of coherent states, with amplitudes
\begin{equation}
\alpha_{L(R)}=\frac{\Omega_{L(R)}}{(-\Delta_{L(R)} \pm gX)-i\kappa/2}
\end{equation}
where $\Omega_{L(R)}$ are related to the incident laser powers by 
$\Omega_{L(R)} = \sqrt{\kappa P_{L(R)}/\hbar \nu_{L(R)}}$,
and $X$ is the steady-state displacement of the membrane, including the fundamental and higher modes.
We note that the fact that the PDH lock
tracks the changes in the cavity frequencies leads to a shift $\nu \rightarrow \nu \mp g X$ for the steady state, effectively enhancing the low frequency contribution to $g$ by a factor of 2 (for details, see the Appendix).

As in the experiment, we only focus on the lowest frequency mechanical mode and symmetric driving and detuning, $\Omega_{L(R)} = \Omega, \Delta_{L(R)} = \Delta$. This mode feels a radiation pressure force and a spring-based restoring force with spring constant $k = m \omega_m^2$, and the steady state is determined by points where the total force is zero and restorative under small variations 
in $X$. Qualitatively, the membrane's motion evolves in a potential combining its internal spring and two competing optical springs.
Zeros in the force (minima and maxima of the potential) occur  when
\begin{equation}
0=kX\left(1-\frac{A}{u^{2}+2u(\kappa^{2}/4-\Delta^{2})+(\Delta^{2}+\kappa^{2}/4)^{2}}\right)\label{e:force}
\end{equation}
where $u= (2 g X)^2$ is the square of the frequency shift including PDH feedback, and the parameter
$A \equiv \frac{-8 \hbar g^2 |\Omega|^2 \Delta}{k}$ is proportional to the pump power. 

This equation has solutions for $u = 0$ and for $u=u_{ss} \equiv \Delta^2 - \kappa^2/4 \pm \sqrt{A-\Delta^2 \kappa^2}$; physical solutions require $u \geq 0$. Implicitly this requires $A > \Delta^2 \kappa^2$, so that $u_{ss}$ is real. 
For $\Delta < \kappa/2$, there is only one non-zero $u$ solution, and the system continuously goes, as a function of power, from $X = 0$ to $X = \pm \frac{\sqrt{u_{ss}}}{2 g}$. This is in direct analogy to a second order buckling transition in a spring, where the broken left/right symmetry is evident even for arbitrarily small values of the displacement.
For $\Delta > \kappa/2$, however, there are two solutions for $u_{ss}$. The smaller corresponds to an unstable branch, while the larger is stable. This leads to a discontinuous change of the membrane displacement at the transition radiation pressure, and provides a first order buckling transition in the optomechanical system. The overall phase diagram, including the experimentally probed regimes, is shown in Fig.~2. Both first and second-order transitions occur, depending upon the detuning, for increasing power. However, at sufficient powers, the system goes unstable (orange region in Fig.~2) and begins to display limit cycle behavior. We generally only consider power levels below this instability.

\begin{figure}
\begin{center}
\includegraphics[width=3in]{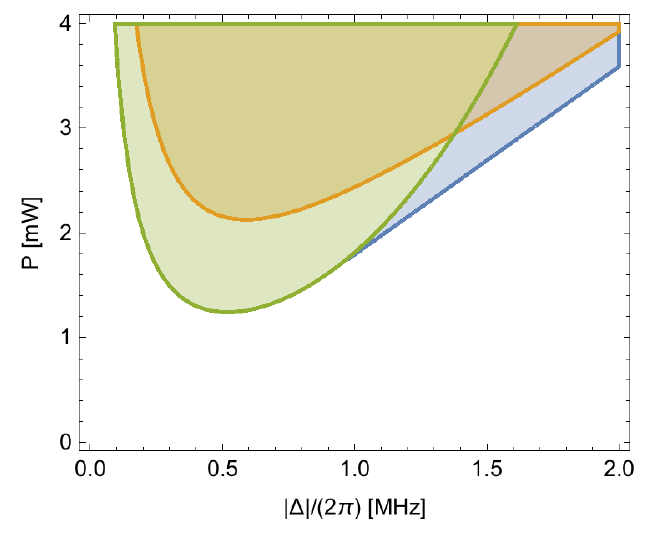}
\caption{Theoretical phase diagram. 
Second order (green) and first order (blue) buckling transitions as a function of laser detuning and power are shown. In addition to the green (second order) and blue (first order) buckled regimes, the nominally unstable region is shown with an orange overlay. 
}
\label{f:theory}
\end{center}
\end{figure}

\begin{figure}
\includegraphics[width=3in]{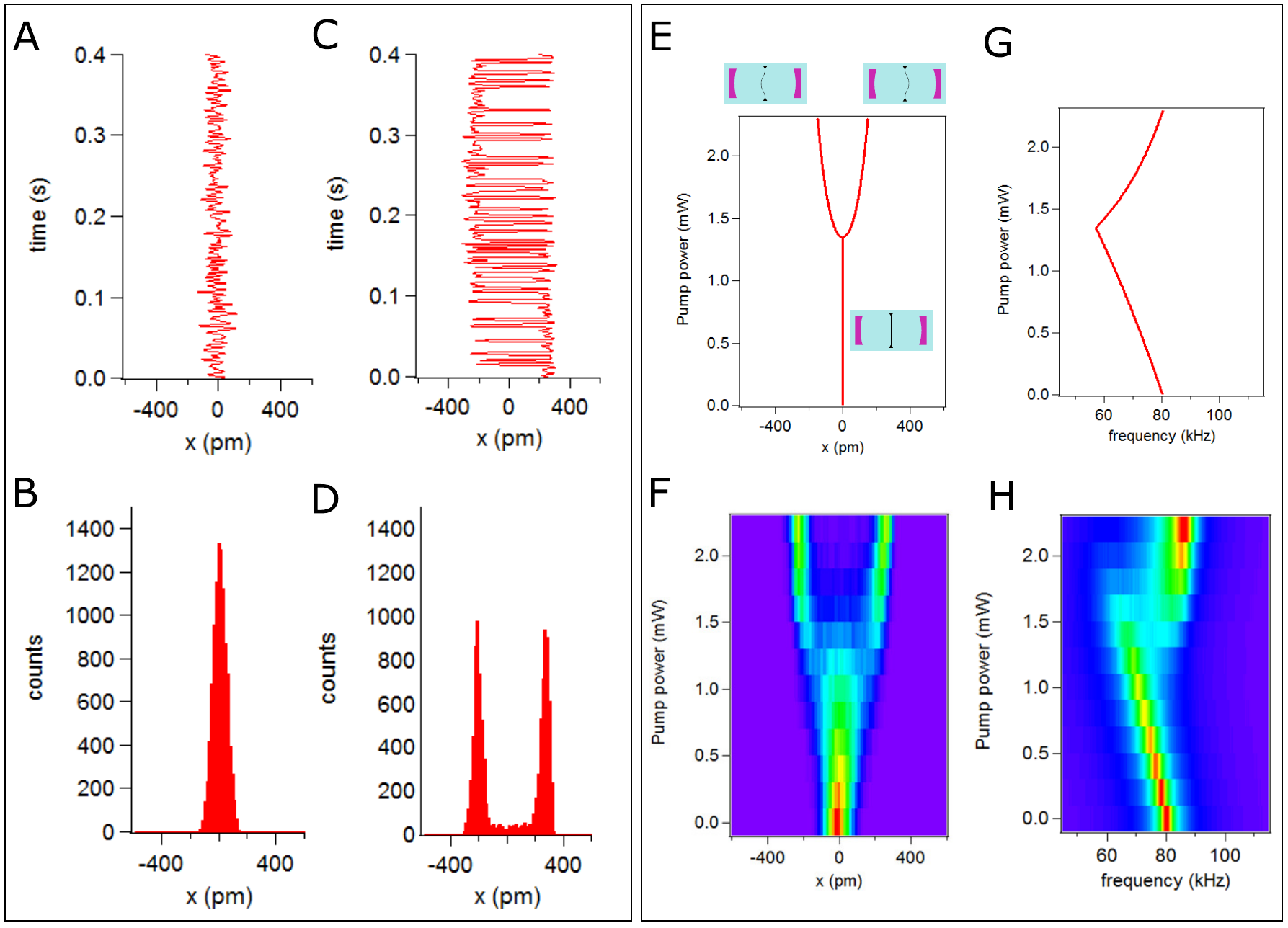}
\caption{\textbf{Buckling transition from single-well to
double-well for \mbox{\boldmath $\Delta=0.22\,\kappa$.} }\newline
\textbf{a,}  Real-time data of the membrane
position without pump lasers. \textbf{b,}  Corresponding histogram. 
The membrane fluctuates around a single position,
resulting from a single-well mechanical potential. \textbf{c,}  Real-time data of the
membrane position with 2.2 mW pump power. \textbf{d,}  Histogram of the membrane
position with 2.2 mW pump power. The membrane fluctuates around two
stable positions, resulting from a double-well optomechanical potential. \textbf{e,}  Calculated
stable positions as a function of pump laser power.  
\textbf{f,}  Image of experimental histograms of the membrane position for increasing pump power. The single-well potential develops smoothly into a double-well potential as the power is raised, showing the onset of the second-order buckling transition.
\textbf{g,}  Calculated mechanical frequency of the membrane for small excursions about the stable positions as a function of pump power.  
\textbf{h,}  Image of mechanical power spectral density inferred from experimental data.  The frequency drops
as the global potential well initially becomes more shallow, then increases as the membrane buckles into
a local potential minimum.}
\end{figure}

We now show the experimental buckling of the optically sprung membrane
for $\Delta=2\pi\times 0.4$~MHz~$=0.22\,\kappa$, where we expect that the dynamics
will correspond to a second order buckling transition. In the absence of a pump laser, the membrane experiences
a pure mechanical single-well quadratic potential and fluctuates around
the stable position due to both thermal and technical noise. 
We record the position of the membrane at a sampling rate
of 2 kHz for 5 seconds; Fig.~3a shows a characteristic subset of the data
for 0.4 seconds, and Fig.~3b shows a histogram of the complete data set,
peaked around zero displacement as expected. When the pump fields are 
turned on, with pump power $P_L=P_R=2.2$~mW, 
the membrane fluctuates around two stable positions, as shown in the time series in Fig.~3c. We attribute
the jumping between stable positions primarily to mechanical noise  in the nanopositioning
stage holding the tethered membrane. As shown in the histogram of the membrane position in Fig.~3d, the membrane buckles to either 
left or right. The steady-state positions predicted by the theory discussed earlier are shown in 
Fig.~3e
as a function of pump laser power for our experimental conditions.  Corresponding experimental
histograms of the membrane position are shown for the same range of pump powers in Fig.~3f.
Both theory and experiment indicate an apparent second order phase transition in the membrane displacement
$X$ as the pump power is raised.  

In addition to the order parameter ($X$), the dynamical response of the system changes
as it passes through the buckling transition.  We observe this by analyzing the spectrum
of the PDH signal for frequencies higher than the bandwidth ($\sim3$~kHz) of the servos used to lock the lasers to the cavity.  Fig.~3h shows that as the pump power is raised, the frequency of the
optically sprung resonator initially diminishes, and then rises above the frequency
of the bare mechanical resonator as the system
buckles.  This is consistent with the picture that the membrane transitions to a double-well 
potential from the sum of the mechanical and optical potentials. 

Curiously, the frequency of the mechanical mode does not go all the way to zero at the phase transition, as might be expected.  This is a consequence of the limited bandwidth of the feedback electronics used to lock the probe lasers to the optical resonances.  Specifically, the opposite frequency dependence with position of the pump lasers and their associated
probes (Fig.~1d) results in a doubling of the optomechanical coupling for displacements
within the bandwidth~($\sim3$~kHz) of the feedback electronics, relative to displacements at substantially
higher frequencies.  Since the mechanical frequency of the membrane is far above this cutoff,
the optical power required to buckle the membrane is well below the power required to drive
its frequency to zero in the unbuckled state.  
Qualitative agreement with the single mechanical mode theory (including the effect of the feedback; see Appendix) is obtained and shown in Fig. 3g, but quantitative agreement likely will require inclusion of the higher mechanical modes whose properties remain challenging to fully characterize in the present setup.

\begin{figure}
\includegraphics[width=3in]{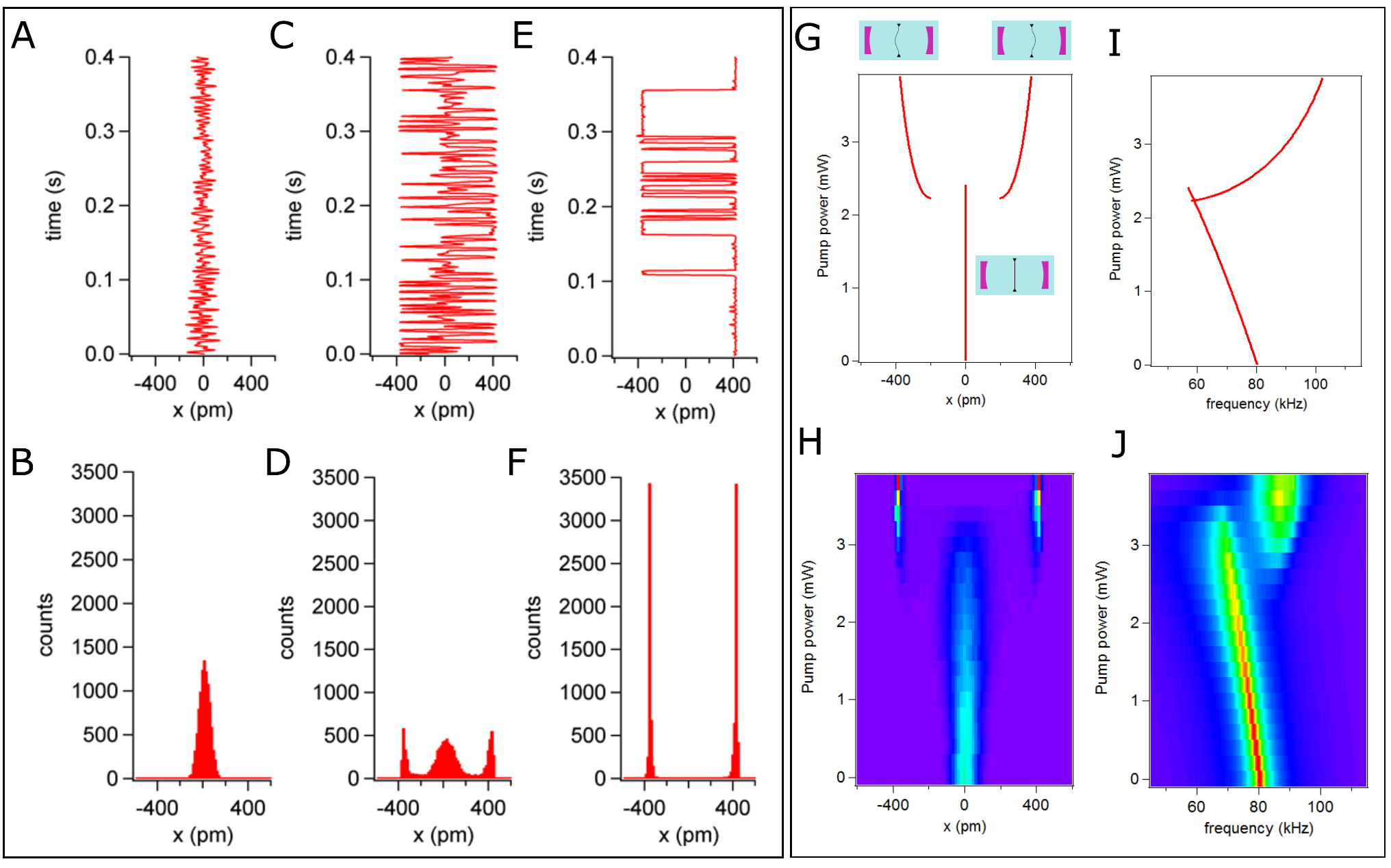}
\caption{\textbf{Buckling transition from single-well to triple-well to
double-well for \mbox{\boldmath $\Delta=0.67\,\kappa$.} }\newline
\textbf{a,} Time series and \textbf{b,} histogram of membrane position,
no pump.  \textbf{c,} Real-time data and 
\textbf{d,} Histogram of the membrane position with 3.0 mW pump power. The membrane fluctuates around three
stable positions, resulting from a triple-well optomechanical potential. \textbf{e,} Real-time data and
\textbf{f,} Histogram of the membrane position with 3.8 mW pump power. The membrane now fluctuates around just two
stable positions, due to a double-well optomechanical potential. 
\textbf{g,} Calculated stable positions as a function of pump laser power.  For a small range of pump powers
there are three stable positions, and as the pump power is raised, the unbuckled state becomes unstable.
\textbf{h,} Image of experimental histograms of the membrane position for increasing pump power. 
\textbf{i,} Calculated mechanical frequency of the membrane for small excursions about the stable positions as a function of pump power.  
\textbf{j,} Image of mechanical power spectral density inferred from experimental data.  
}
\end{figure}

Based upon the theoretical understanding of the phase diagram, we
expect the buckling transition to be qualitatively different for detunings
$\Delta\ge\kappa/2$. Fig.~4 shows our examination of the buckling transition for $\Delta=2\pi\times 1.2$~MHz~$=0.67\,\kappa$.
Once again, Fig.~4a and Fig.~4b depict the noise-induced fluctuations of the membrane in
the absence of pump lasers. Fig.~4c and Fig.~4d show the time series and histograms of
membrane position for $P_L=P_R=3.0$~mW; this time, the membrane fluctuates around three stable positions, 
either remaining unbuckled or buckling to either left or right.  When the pump powers are raised to 3.8~mW, only the buckled states remain stable, as shown in Fig.~4e and Fig.~4f.

Theoretical predictions of the steady states as a function of pump power are shown for our
experimental conditions in Fig.~4g, and the corresponding experimental
histograms of the membrane position are shown in Fig.~4h.  In addition to the initial unbuckled state,
two more stable positions appear discontinuously as the pump power is increased,  
indicating that the membrane now experiences an effective triple-well
potential. This jump to a finite displacement of the membrane corresponds to a nonequilibrium first order phase transition.  As the power is raised still further, the steady state at zero displacement
becomes unstable, and the potential becomes a double well. 

Examination of the mechanical oscillation frequency, Fig.~4j, once again reveals 
that the frequency of the optically sprung oscillator initially diminishes with pump power, but 
jumps to a value larger than the bare mechanical frequency in the final double-well regime.
For a small range of powers, corresponding to the triple-well regime, the frequency distribution
is bimodal.  The corresponding theoretical curve, including feedback as discussed previously, is shown in
Fig.~4i, where it is clear that different mechanical frequencies are expected in the local
potential minima corresponding to the buckled and unbuckled states in the triple-well potential regime.

We note that the power levels chosen push into the nominally unstable region of the phase diagram for the large detuning data. 
The initial behavior in this unstable region, however, corresponds to a limit cycle behavior analogous to that of a Duffing oscillator, leading to deviations in the experiment from the simple theoretical picture presented earlier, as the system wanders in a potential landscape with position-dependent gain and loss. 
We believe that this fact, coupled with the feedback and the restriction of our theoretical model
to a single mechanical mode, are responsible for 
the differences in the shapes of the experimental data
in Fig.~4h and Fig.~4j from their theoretical counterparts Fig.~4g and Fig.~4i, rather than
any essential physics not described already.

Our system is in the classical regime, working at room temperature and with low quality factor. However, at low temperatures and high quality factor, a quantum phase transition may be observable in systems of this nature\cite{mumford_dicke-type_2015}. Specifically, optomechanical systems can be made sufficiently cold -- with a nominal dephasing rate slower than their resonance frequency -- and sideband resolved to be laser cooled to their groundstate before buckling\cite{chan_laser_2011}. Then a rapid increase in pump power bringing the system across the transition could yield a transition driven entirely by quantum fluctuations, a macroscopic version of structural quantum phase transitions such as those in ion crystals\cite{Shimshoni2011}.

In conclusion, we have realized an optomechanical system in which it is possible to force mechanical 
(buckling) phase transitions of the first or second kind, using the detuning as the control parameter
and the optical power to effect the transition.  It is straightforward to understand the 
physics in terms of a configurable optical spring creating double-well and triple-well mechanical
potentials.  Furthermore, we have presented an analytic theory with remarkably simple results
that captures nearly all of the experimental results that we observe.  This approach provides a new platform for
further studying quantum physics such as macroscopic quantum tunneling\cite{buchmann_macroscopic_2012} when taken to the cryogenic domain. In addition, our results suggest a variety of potential applications to active and passive optical devices, including low-power switches, power filters, and  self-excited oscillators.

We acknowledge helpful discussions with J. Harris, A. Schliesser, and E. Polzik. Funding was provided by DARPA QuASAR and the NSF-funded Physics Frontier Center at the
JQI. Research performed in part at the NIST Center for Nanoscale Science and Technology.

\appendix
\section{Derivation of the phase diagram}
 Consider a two-mode optical system where the optical frequency difference is much larger than any mechanical scale in the problem. We can then neglect beating between the two optical fields, and work with each cavity mode rotating in a frame near resonant with the closer of the two pump lasers with Rabi frequencies $\Omega_L, \Omega_R$. Labeling the modes L and R (where L represents a mode that increases in frequency as the mechanical oscillator moves leftward, and R the opposite), we have equations of motion for the fields $a_L, a_R$ in this rotating frame:
\begin{align}
\dot{a}_L &= i (\Delta_L - g_L x) a_L - \frac{\kappa_L}{2} a_L + \sqrt{\kappa_L} a_{L,\textrm{in}} + i \Omega_L \label{e:L} \\
\dot{a}_R &= i (\Delta_R + g_R x) a_R - \frac{\kappa_R}{2} a_R + \sqrt{\kappa_R} a_{R,\textrm{in}} + i \Omega_R \label{e:R} 
\end{align}
where $\Delta_L = \nu_L - \omega_L$ is the detuning between the $L$ pump laser and the cavity $L$ mode (initially negative for our setup), $a_{L(R),\textrm{in}}$ are the input vacuum fields neglected in the classical analysis that follows, and $x$ is the operator representing the displacement of the mechanical oscillator along the cavity axis relative to the zero pump power position, which in general involves the motion of multiple mechanical modes. The coupling constants $g_L, g_R$ are measured in the experiment by observing the change of angular frequency for the cavity as a function of position of the membrane. We neglect terms beating at the laser frequency difference ($\sim 8$ GHz, and thus faster than any other scale in the problem). In the experimental setup, $\kappa_L \sim \kappa_R = \kappa$. The radiation pressure force is $F_{\textrm{rad}} \approx -\hbar (g_L a_L^\dag a_L - g_R a_R^\dag a_R)$. Note that for our system $g_L = g_R = g$, and we similarly set $\kappa_L = \kappa_R = \kappa$ for simplicity, neglecting potential dispersive effects.

When an optical system is pumped with a laser system, the behavior is best described by finding the classical steady state of the system and looking at fluctuations around that steady state. For simplicity, we start with the scenario where only the fundamental mechanical mode with spring constant $k$ is coupled to the optical system. We can look for the steady state of the driven system by solving the  equations for coherent state amplitudes $\alpha_{L},\alpha_{R}$. This yields
\begin{equation}
\alpha_{L(R)} = \frac{\Omega_{L(R)}}{(-\Delta_{L(R)} \pm g X) - i \kappa/2} 
\end{equation} 
where $X$ is the steady-state position of the mechanical resonator.

To go from the cavity amplitudes to the full steady state requires examining both the cavity and the mechanical equations of motion.  In practice, our description must also include the action of the feedback electronics, particularly if we want to know about the dynamics of our system. Fortunately, inclusion of the feedback can be included with a simple model with an additional degree of freedom $\delta$ representing the action of laser locking by the probe beams, which leads to slow feedback in the detunings of the pump fields. The equations become
\begin{align}
\dot{x} &= p/m \\
\dot{p} &= -k x - \gamma_m p + F_{\textrm{rad}} \\
\dot{\delta} & = - \frac{1}{\tau_{FB}} (\delta - g x) \\
\Delta_L &= \Delta_0 - \delta \\
\Delta_R &= \Delta_0 + \delta
\label{e:eom}
\end{align}
where $\tau_{FB} \gg \sqrt{k/m}$ is the feedback timescale, around 100 $\mu$s in the experiment, and $\Delta_0$ is the bare detuning offset set by the EOMs. 

On very slow time scales, the system goes to a steady state with $\delta_{ss} = g X$ (enabling readout of the position of the resonator using the feedback circuit). In practice, the effect of the feedback will be to double the optomechanical response in the system. The oscillator position is found by solving for the zero force condition 
\begin{equation}
k X = - \hbar g (|\alpha_L|^2 - |\alpha_R|^2)
\end{equation}

For the symmetric case, finding the steady state corresponds to solving a surprising simple equation for zero force.  The zero force condition becomes
\begin{widetext}
\begin{equation}
0 = k X \left(1 + \frac{8 \hbar g^2 \Delta_0 \Omega^2}{k} \frac{1}{\kappa^4/16 + (4 g^2 X^2 + \Delta_0^2) \kappa^2/2 + (\Delta_0^2 - 4 g^2 X^2)^2} \right)
\end{equation}
\end{widetext}
This has the trivial (low power) solution $X = 0$. In addition, at higher power it admits additional solutions.

To examine the higher power solutions, we define $u \equiv (2 g X)^2$ as a generalized position variable and $A \equiv -8 \hbar g^2 |\Omega|^2 \Delta_0 / k$ as a variable proportional to the incoming optical power. Note that $\Delta_0 < 0$ (red detuning) for the experiment, and thus $A>0$. The zero force condition has solutions in $u$ according to the quadratic equation:
\begin{equation}
u_{ss,\pm} =  \Delta_0^2 - \kappa^2/4 \pm \sqrt{A - \Delta_0^2 \kappa^2} 
\end{equation}
Solutions to these equations provide the phase diagram shown in the main text.

We examine which of these solutions is physical. We require $X$ to be real, or equivalently $u > 0$ and real. If $|\Delta_0|  < \kappa/2$, only the positive solution $u_{ss,+}$, can satisfy $u>0$. At the critical value of $A = A_{2} \equiv (\Delta_0^2 + \kappa^2/4)^2$, $u_{ss,+} = 0$, and above this, $u$ continuously takes a non-zero value. This corresponds to the second-order phase transition as described in the main text, and can be understood as a double well potential for $X$

When $|\Delta_0| > \kappa/2$, both branches may satisfy $u>0$.  The requirement of $u$ real is equivalent to $A > A_{1} \equiv \Delta_0^2 \kappa^2$, at which $u$ discontinuously takes a non-zero value. This corresponds to the first-order phase transition as described in the main text.  This  second regime corresponds to a triple well potential for $X$, with the smaller solutions for $u \neq 0$  unstable (the peaks of the barriers between wells). Thus, if we conceive of a non equilibrium phase diagram in which optical power is varied, for small detuning a second order phase transition will occur, while for larger detunings, a first order (discontinuous change of $X$ as a function of $A$) transition occurs.

One potential limit to stability for these systems is the detuning of the cavity modes becoming sufficiently modified by the transition to go from `red' to `blue'. This occurs when $2 g |X| > |\Delta_0|$ or $u > \Delta_0^2$, which occurs for
\beq
A > A_{\textrm{inst}} \equiv \kappa^4/16 + \Delta_0^2 \kappa^2\ .
\eeq
In practice, the stability of the overall system can be maintained even when one of the cavity modes is blue detuned, given sufficient mechanical damping. In addition, we find via numerical simulations that larger values of $A$ lead to limit cycle behavior that behaves similarly to the steady state solutions already found.


\section{Dynamical response}

In addition to the steady state solutions, we can examine the dynamical response of the system near its steady state. As our system is operating in the limit of cavity linewidth much greater than mechanical frequency, we anticipate that the dominant corrections to the bare resonator behavior take the form of the optical spring effect. 

Formally, we can find this behavior via expansion of the equations of motion for small excursions from the steady state solutions. Specifically, for equations of the motion of the form $\dot v_\mu = M_\mu(\vec{v}) - F_\mu$, such as those given in Eqs.~1-2,4-6, we find the steady state $\vec{v}$ and expand around it with $\delta \vec{v}$. The corresponding equations of motion are
\[
\dot \delta v_\mu = (\partial_\nu M_\mu |_{\vec{v}} ) \delta v_\nu
\]
where the Einstein summation condition is implied for Greek indices. Moving to the Fourier domain with frequency coordinate $\nu$, we can eliminate the equations involving fluctuations of the cavity fields, finding they directly depend upon the position fluctuations $x$ and the detuning fluctuations $\delta$. 

Taking $\tau_{FB} \gg 1/\nu$ but $\nu \ll \kappa$, we can expand the corrections from the cavity coordinates to recover the optical spring result: the spring constant $k$ is modified, as is the mechanical damping $\gamma$ and the effective inertial mass $m$:
\begin{widetext}
\begin{align}
k_{\textrm{eff}} & = k +  2 g^2 \hbar \left( \frac{|\alpha_L|^2 (\Delta_0 - 2 g X)}{(\Delta_0 - 2 g X)^2 + \kappa^2/4} + \frac{|\alpha_R|^2 (\Delta_0 + 2 g X)}{(\Delta_0 + 2 g X)^2 + \kappa^2/4} \right) \\
\gamma_{\textrm{eff}} &= \gamma + \frac{2}{m} \partial_\kappa k_{\textrm{eff}} \\
m_{\textrm{eff}} &= m + 2 \partial^2_{\kappa} k_{\textrm{eff}}
\end{align}
\end{widetext}
where partial derivatives assume that steady state values $\alpha_L,\alpha_R,X$ are independent of $\kappa$. In practice, for our parameters, the difference in the effective mass is negligible. These formula are then used to find the dynamical response and the stable region of the phase diagram given in the main text.  In contrast to the case without feedback, here feedback causes the mechanical mode to have a non-zero frequency at the phase transition.

%

While qualitatively the behavior is consistent between the theory and experiment, several simplifying assumptions preclude quantitative agreement. Our specific concerns include the role high frequency mechanical modes play. In addition, with the high circulating power in our system, substantial heating of the tethered membrane is expected.

\begin{figure*}
\includegraphics[width=7in]{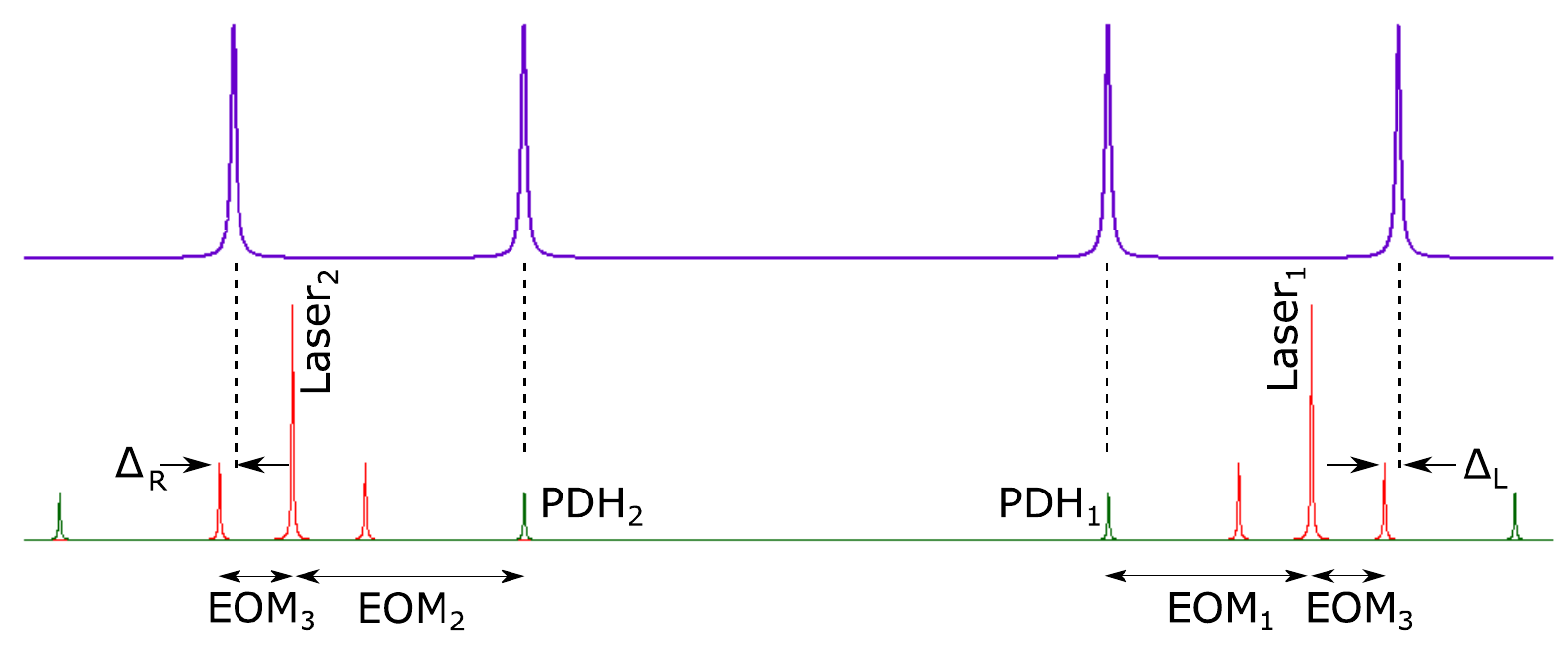}
\caption{
The relevant cavity modes (top) and spectrum showing how pump and probe laser fields
are generated with electro-optic phase modulators (bottom).  Laser fields that are
not near-resonant with cavity modes are rejected by the cavity and play no role.
Sidebands generated by $\mbox{EOM}_{4}$ and $\mbox{EOM}_{5}$ to obtain a
Pound-Drever-Hall error signal are not shown in the interest of clarity. }
\end{figure*}

\section{Generation of pump and probe fields}
We clarify here how the pump and probe fields represented in Fig.~1d of the main text are
generated.  $\mbox{Laser}_{1}$ and $\mbox{Laser}_{2}$ are independent tunable lasers with $\lambda=1560$~nm. 
Referring to Fig.~5 of the Appendix, electro-optic phase modulators $\mbox{EOM}_{1}$ and $\mbox{EOM}_{2}$ are driven
independently with frequencies in the vicinity of 1.5~GHz, and generate sidebands
on $\mbox{Laser}_{1}$ and $\mbox{Laser}_{2}$.  The lower-frequency sideband of $\mbox{EOM}_{1}$
is denoted PDH$_1$ and locked to one mode of the cavity, and the upper-frequency sideband of $\mbox{EOM}_{2}$
is denoted PDH$_2$ and locked to an adjacent mode.  Locking is accomplished by the Pound-Drever-Hall (PDH)
method, using electro-optic modulators $\mbox{EOM}_{4}$ and $\mbox{EOM}_{5}$ to generate sidebands (not shown
in the figure) at 20~MHz.  Part of the light from
$\mbox{Laser}_{1}$ and $\mbox{Laser}_{2}$ is combined and amplified in an erbium-doped fiber
amplifier and passed through phase modulator $\mbox{EOM}_{3}$, driven at a frequency of approximately 0.5~GHz.  The upper-frequency sideband 
of $\mbox{Laser}_{1}$ is denoted Pump$_1$ and the lower-frequency sideband 
of $\mbox{Laser}_{2}$ is denoted Pump$_2$. The detunings $\Delta_L$ and $\Delta_R$ (taken to be equal in the experiment) of the pump fields from their associated cavity
modes may be controlled independently by choice of the frequencies driving
$\mbox{EOM}_{1}$ and $\mbox{EOM}_{2}$.  The probe fields are combined and sent 
into one port of the cavity, and the pump beams are combined and sent into the other port.

\section{Fabrication of tethered membrane}

We start with a double side polished silicon wafer as the substrate.
First we deposit a layer of silicon nitride film on both sides of
the wafer using a low pressure chemical vapor deposition furnace. The
thickness of the silicon nitride film is measured to be about 258 nm by a spectroscopic
ellipsometer, limited by the instrumental resolution. Then
we do photolithography and plasma etching to open a square window
in the silicon nitride film on one side of the wafer. Removing the
silicon in the window by wet etching using potassium hydroxide solution,
we obtain a square suspended silicon nitride membrane. The membrane
is furthered shaped to a tethered membrane ($200\:\mu\mbox{m}\times200\:\mu\mbox{m}$)
with long thin tethers ($2\:\mu\mbox{m}$ wide) connecting the membrane
to its frame using electron-beam lithography and plasma etching. This is done to reduce the effective spring constant and allow for a substantial optical spring effect at low
laser power. Finite element analysis is used to simulate the mechanical properties of the
tethered membrane to aid in the design.

\section{Position readout of the tethered membrane}
We use two different schemes for the measurement of small membrane
position variations. For measurements in a bandwidth below 1~kHz,
we employ the fact that the variation of the beat frequency between
the probe lasers is locally proportional to the variation of the membrane
position. Since the probe fields contain additional frequency components 
generated by EOMs, we instead measure the beat frequency between the (locked) original lasers, whose
frequencies directly track those of the probes.  
The beat frequency measurement is performed with a high-frequency counter. We
use this approach for low-frequency measurements, such as in measuring the stable position of the membrane before and after the buckling transitions. 

For measuring mechanical motion at frequencies substantially higher than the PDH servo
bandwidth ($\sim3$~kHz), we use the PDH signal as a probe of membrane position variation. The power spectral densities shown in Fig.~3h and Fig.~4j are obtained by Fourier transform of the PDH signal.


\bibliographystyle{apsrev}

\begin{thebibliography}{30}
\expandafter\ifx\csname natexlab\endcsname\relax\def\natexlab#1{#1}\fi
\expandafter\ifx\csname bibnamefont\endcsname\relax
  \def\bibnamefont#1{#1}\fi
\expandafter\ifx\csname bibfnamefont\endcsname\relax
  \def\bibfnamefont#1{#1}\fi
\expandafter\ifx\csname citenamefont\endcsname\relax
  \def\citenamefont#1{#1}\fi
\expandafter\ifx\csname url\endcsname\relax
  \def\url#1{\texttt{#1}}\fi
\expandafter\ifx\csname urlprefix\endcsname\relax\def\urlprefix{URL }\fi
\providecommand{\bibinfo}[2]{#2}
\providecommand{\eprint}[2][]{\url{#2}}

\bibitem[{\citenamefont{Dorsel et~al.}(1983)\citenamefont{Dorsel, McCullen,
  Meystre, Vignes, and Walther}}]{dorsel_optical_1983}
\bibinfo{author}{\bibfnamefont{A.}~\bibnamefont{Dorsel}},
  \bibinfo{author}{\bibfnamefont{J.~D.} \bibnamefont{McCullen}},
  \bibinfo{author}{\bibfnamefont{P.}~\bibnamefont{Meystre}},
  \bibinfo{author}{\bibfnamefont{E.}~\bibnamefont{Vignes}}, \bibnamefont{and}
  \bibinfo{author}{\bibfnamefont{H.}~\bibnamefont{Walther}},
  \bibinfo{journal}{Physical Review Letters} \textbf{\bibinfo{volume}{51}},
  \bibinfo{pages}{1550} (\bibinfo{year}{1983}),
  \urlprefix\url{http://journals.aps.org/prl/abstract/10.1103/PhysRevLett.51.1550}.

\bibitem[{\citenamefont{Meystre et~al.}(1985)\citenamefont{Meystre, Wright,
  McCullen, and Vignes}}]{meystre_theory_1985}
\bibinfo{author}{\bibfnamefont{P.}~\bibnamefont{Meystre}},
  \bibinfo{author}{\bibfnamefont{E.~M.} \bibnamefont{Wright}},
  \bibinfo{author}{\bibfnamefont{J.~D.} \bibnamefont{McCullen}},
  \bibnamefont{and} \bibinfo{author}{\bibfnamefont{E.}~\bibnamefont{Vignes}},
  \bibinfo{journal}{JOSA B} \textbf{\bibinfo{volume}{2}}, \bibinfo{pages}{1830}
  (\bibinfo{year}{1985}),
  \urlprefix\url{http://www.opticsinfobase.org/abstract.cfm?uri=josab-2-11-1830}.

\bibitem[{\citenamefont{Milburn and Woolley}(2011)}]{milburn_introduction_2011}
\bibinfo{author}{\bibfnamefont{G.}~\bibnamefont{Milburn}} \bibnamefont{and}
  \bibinfo{author}{\bibfnamefont{M.}~\bibnamefont{Woolley}},
  \bibinfo{journal}{Acta Physica Slovaca. Reviews and Tutorials}
  \textbf{\bibinfo{volume}{61}}, \bibinfo{pages}{483} (\bibinfo{year}{2011}),
  ISSN \bibinfo{issn}{1336-040X, 0323-0465},
  \urlprefix\url{http://www.degruyter.com/view/j/apsrt.2011.61.issue-5/v10155-011-0005-7/v10155-011-0005-7.xml}.

\bibitem[{\citenamefont{Meystre}(2013)}]{meystre_short_2013}
\bibinfo{author}{\bibfnamefont{P.}~\bibnamefont{Meystre}},
  \bibinfo{journal}{Annalen der Physik} \textbf{\bibinfo{volume}{525}},
  \bibinfo{pages}{215} (\bibinfo{year}{2013}), ISSN \bibinfo{issn}{00033804},
  \urlprefix\url{http://doi.wiley.com/10.1002/andp.201200226}.

\bibitem[{\citenamefont{Aspelmeyer et~al.}(2014)\citenamefont{Aspelmeyer,
  Kippenberg, and Marquardt}}]{aspelmeyer_cavity_2014}
\bibinfo{author}{\bibfnamefont{M.}~\bibnamefont{Aspelmeyer}},
  \bibinfo{author}{\bibfnamefont{T.~J.} \bibnamefont{Kippenberg}},
  \bibnamefont{and}
  \bibinfo{author}{\bibfnamefont{F.}~\bibnamefont{Marquardt}},
  \bibinfo{journal}{Reviews of Modern Physics} \textbf{\bibinfo{volume}{86}},
  \bibinfo{pages}{1391} (\bibinfo{year}{2014}), ISSN \bibinfo{issn}{0034-6861,
  1539-0756},
  \urlprefix\url{http://link.aps.org/doi/10.1103/RevModPhys.86.1391}.

\bibitem[{\citenamefont{Metcalfe}(2014)}]{metcalfe_applications_2014}
\bibinfo{author}{\bibfnamefont{M.}~\bibnamefont{Metcalfe}},
  \bibinfo{journal}{Applied Physics Reviews} \textbf{\bibinfo{volume}{1}},
  \bibinfo{pages}{031105} (\bibinfo{year}{2014}), ISSN
  \bibinfo{issn}{1931-9401},
  \urlprefix\url{http://scitation.aip.org/content/aip/journal/apr2/1/3/10.1063/1.4896029}.

\bibitem[{\citenamefont{Abramovici et~al.}(1992)\citenamefont{Abramovici,
  Althouse, Drever, Gursel, Kawamura, Raab, Shoemaker, Sievers, Spero, Thorne
  et~al.}}]{abramovici_ligo:_1992}
\bibinfo{author}{\bibfnamefont{A.}~\bibnamefont{Abramovici}},
  \bibinfo{author}{\bibfnamefont{W.~E.} \bibnamefont{Althouse}},
  \bibinfo{author}{\bibfnamefont{R.~W.~P.} \bibnamefont{Drever}},
  \bibinfo{author}{\bibfnamefont{Y.}~\bibnamefont{Gursel}},
  \bibinfo{author}{\bibfnamefont{S.}~\bibnamefont{Kawamura}},
  \bibinfo{author}{\bibfnamefont{F.~J.} \bibnamefont{Raab}},
  \bibinfo{author}{\bibfnamefont{D.}~\bibnamefont{Shoemaker}},
  \bibinfo{author}{\bibfnamefont{L.}~\bibnamefont{Sievers}},
  \bibinfo{author}{\bibfnamefont{R.~E.} \bibnamefont{Spero}},
  \bibinfo{author}{\bibfnamefont{K.~S.} \bibnamefont{Thorne}},
  \bibnamefont{et~al.}, \bibinfo{journal}{Science}
  \textbf{\bibinfo{volume}{256}}, \bibinfo{pages}{325} (\bibinfo{year}{1992}),
  ISSN \bibinfo{issn}{0036-8075, 1095-9203},
  \urlprefix\url{http://www.sciencemag.org/cgi/doi/10.1126/science.256.5055.325}.

\bibitem[{\citenamefont{Li et~al.}(2008)\citenamefont{Li, Pernice, Xiong,
  Baehr-Jones, Hochberg, and Tang}}]{li_harnessing_2008}
\bibinfo{author}{\bibfnamefont{M.}~\bibnamefont{Li}},
  \bibinfo{author}{\bibfnamefont{W.~H.~P.} \bibnamefont{Pernice}},
  \bibinfo{author}{\bibfnamefont{C.}~\bibnamefont{Xiong}},
  \bibinfo{author}{\bibfnamefont{T.}~\bibnamefont{Baehr-Jones}},
  \bibinfo{author}{\bibfnamefont{M.}~\bibnamefont{Hochberg}}, \bibnamefont{and}
  \bibinfo{author}{\bibfnamefont{H.~X.} \bibnamefont{Tang}},
  \bibinfo{journal}{Nature} \textbf{\bibinfo{volume}{456}},
  \bibinfo{pages}{480} (\bibinfo{year}{2008}), ISSN \bibinfo{issn}{0028-0836,
  1476-4687},
  \urlprefix\url{http://www.nature.com/doifinder/10.1038/nature07545}.

\bibitem[{\citenamefont{Ludwig et~al.}(2012)\citenamefont{Ludwig,
  Safavi-Naeini, Painter, and Marquardt}}]{ludwig_enhanced_2012}
\bibinfo{author}{\bibfnamefont{M.}~\bibnamefont{Ludwig}},
  \bibinfo{author}{\bibfnamefont{A.~H.} \bibnamefont{Safavi-Naeini}},
  \bibinfo{author}{\bibfnamefont{O.}~\bibnamefont{Painter}}, \bibnamefont{and}
  \bibinfo{author}{\bibfnamefont{F.}~\bibnamefont{Marquardt}},
  \bibinfo{journal}{Physical Review Letters} \textbf{\bibinfo{volume}{109}},
  \bibinfo{pages}{063601} (\bibinfo{year}{2012}), ISSN
  \bibinfo{issn}{0031-9007, 1079-7114},
  \urlprefix\url{http://link.aps.org/doi/10.1103/PhysRevLett.109.063601}.

\bibitem[{\citenamefont{Winger et~al.}(2011)\citenamefont{Winger, Blasius,
  Mayer~Alegre, Safavi-Naeini, Meenehan, Cohen, Stobbe, and
  Painter}}]{winger_chip-scale_2011}
\bibinfo{author}{\bibfnamefont{M.}~\bibnamefont{Winger}},
  \bibinfo{author}{\bibfnamefont{T.~D.} \bibnamefont{Blasius}},
  \bibinfo{author}{\bibfnamefont{T.~P.} \bibnamefont{Mayer~Alegre}},
  \bibinfo{author}{\bibfnamefont{A.~H.} \bibnamefont{Safavi-Naeini}},
  \bibinfo{author}{\bibfnamefont{S.}~\bibnamefont{Meenehan}},
  \bibinfo{author}{\bibfnamefont{J.}~\bibnamefont{Cohen}},
  \bibinfo{author}{\bibfnamefont{S.}~\bibnamefont{Stobbe}}, \bibnamefont{and}
  \bibinfo{author}{\bibfnamefont{O.}~\bibnamefont{Painter}},
  \bibinfo{journal}{Optics Express} \textbf{\bibinfo{volume}{19}},
  \bibinfo{pages}{24905} (\bibinfo{year}{2011}), ISSN
  \bibinfo{issn}{1094-4087},
  \urlprefix\url{http://www.opticsinfobase.org/abstract.cfm?URI=oe-19-25-24905}.

\bibitem[{\citenamefont{Krause et~al.}(2012)\citenamefont{Krause, Winger,
  Blasius, Lin, and Painter}}]{krause_high-resolution_2012}
\bibinfo{author}{\bibfnamefont{A.~G.} \bibnamefont{Krause}},
  \bibinfo{author}{\bibfnamefont{M.}~\bibnamefont{Winger}},
  \bibinfo{author}{\bibfnamefont{T.~D.} \bibnamefont{Blasius}},
  \bibinfo{author}{\bibfnamefont{Q.}~\bibnamefont{Lin}}, \bibnamefont{and}
  \bibinfo{author}{\bibfnamefont{O.}~\bibnamefont{Painter}},
  \bibinfo{journal}{Nature Photonics} \textbf{\bibinfo{volume}{6}},
  \bibinfo{pages}{768} (\bibinfo{year}{2012}), ISSN \bibinfo{issn}{1749-4885,
  1749-4893},
  \urlprefix\url{http://www.nature.com/doifinder/10.1038/nphoton.2012.245}.

\bibitem[{\citenamefont{Sheard et~al.}(2004)\citenamefont{Sheard, Gray,
  Mow-Lowry, McClelland, and Whitcomb}}]{sheard_observation_2004}
\bibinfo{author}{\bibfnamefont{B.~S.} \bibnamefont{Sheard}},
  \bibinfo{author}{\bibfnamefont{M.~B.} \bibnamefont{Gray}},
  \bibinfo{author}{\bibfnamefont{C.~M.} \bibnamefont{Mow-Lowry}},
  \bibinfo{author}{\bibfnamefont{D.~E.} \bibnamefont{McClelland}},
  \bibnamefont{and} \bibinfo{author}{\bibfnamefont{S.~E.}
  \bibnamefont{Whitcomb}}, \bibinfo{journal}{Physical Review A}
  \textbf{\bibinfo{volume}{69}}, \bibinfo{pages}{051801(R)}
  (\bibinfo{year}{2004}), ISSN \bibinfo{issn}{1050-2947, 1094-1622},
  \urlprefix\url{http://link.aps.org/doi/10.1103/PhysRevA.69.051801}.

\bibitem[{\citenamefont{Di~Virgilio et~al.}(2006)\citenamefont{Di~Virgilio,
  Barsotti, Braccini, Bradaschia, Cella, Corda, Dattilo, Ferrante, Fidecaro,
  Fiori et~al.}}]{VirgoOptSpring2006}
\bibinfo{author}{\bibfnamefont{A.}~\bibnamefont{Di~Virgilio}},
  \bibinfo{author}{\bibfnamefont{L.}~\bibnamefont{Barsotti}},
  \bibinfo{author}{\bibfnamefont{S.}~\bibnamefont{Braccini}},
  \bibinfo{author}{\bibfnamefont{C.}~\bibnamefont{Bradaschia}},
  \bibinfo{author}{\bibfnamefont{G.}~\bibnamefont{Cella}},
  \bibinfo{author}{\bibfnamefont{C.}~\bibnamefont{Corda}},
  \bibinfo{author}{\bibfnamefont{V.}~\bibnamefont{Dattilo}},
  \bibinfo{author}{\bibfnamefont{I.}~\bibnamefont{Ferrante}},
  \bibinfo{author}{\bibfnamefont{F.}~\bibnamefont{Fidecaro}},
  \bibinfo{author}{\bibfnamefont{I.}~\bibnamefont{Fiori}},
  \bibnamefont{et~al.}, \bibinfo{journal}{Physical Review A}
  \textbf{\bibinfo{volume}{74}}, \bibinfo{pages}{013813}
  (\bibinfo{year}{2006}), ISSN \bibinfo{issn}{1050-2947, 1094-1622},
  \urlprefix\url{http://link.aps.org/doi/10.1103/PhysRevA.74.013813}.

\bibitem[{\citenamefont{Corbitt
  et~al.}(2007{\natexlab{a}})\citenamefont{Corbitt, Chen, Innerhofer,
  Müller-Ebhardt, Ottaway, Rehbein, Sigg, Whitcomb, Wipf, and
  Mavalvala}}]{corbitt_all-optical_2007}
\bibinfo{author}{\bibfnamefont{T.}~\bibnamefont{Corbitt}},
  \bibinfo{author}{\bibfnamefont{Y.}~\bibnamefont{Chen}},
  \bibinfo{author}{\bibfnamefont{E.}~\bibnamefont{Innerhofer}},
  \bibinfo{author}{\bibfnamefont{H.}~\bibnamefont{Müller-Ebhardt}},
  \bibinfo{author}{\bibfnamefont{D.}~\bibnamefont{Ottaway}},
  \bibinfo{author}{\bibfnamefont{H.}~\bibnamefont{Rehbein}},
  \bibinfo{author}{\bibfnamefont{D.}~\bibnamefont{Sigg}},
  \bibinfo{author}{\bibfnamefont{S.}~\bibnamefont{Whitcomb}},
  \bibinfo{author}{\bibfnamefont{C.}~\bibnamefont{Wipf}}, \bibnamefont{and}
  \bibinfo{author}{\bibfnamefont{N.}~\bibnamefont{Mavalvala}},
  \bibinfo{journal}{Physical Review Letters} \textbf{\bibinfo{volume}{98}},
  \bibinfo{pages}{150802} (\bibinfo{year}{2007}{\natexlab{a}}), ISSN
  \bibinfo{issn}{0031-9007, 1079-7114},
  \urlprefix\url{http://link.aps.org/doi/10.1103/PhysRevLett.98.150802}.

\bibitem[{\citenamefont{Corbitt
  et~al.}(2007{\natexlab{b}})\citenamefont{Corbitt, Wipf, Bodiya, Ottaway,
  Sigg, Smith, Whitcomb, and Mavalvala}}]{corbitt_optical_2007}
\bibinfo{author}{\bibfnamefont{T.}~\bibnamefont{Corbitt}},
  \bibinfo{author}{\bibfnamefont{C.}~\bibnamefont{Wipf}},
  \bibinfo{author}{\bibfnamefont{T.}~\bibnamefont{Bodiya}},
  \bibinfo{author}{\bibfnamefont{D.}~\bibnamefont{Ottaway}},
  \bibinfo{author}{\bibfnamefont{D.}~\bibnamefont{Sigg}},
  \bibinfo{author}{\bibfnamefont{N.}~\bibnamefont{Smith}},
  \bibinfo{author}{\bibfnamefont{S.}~\bibnamefont{Whitcomb}}, \bibnamefont{and}
  \bibinfo{author}{\bibfnamefont{N.}~\bibnamefont{Mavalvala}},
  \bibinfo{journal}{Physical Review Letters} \textbf{\bibinfo{volume}{99}},
  \bibinfo{pages}{160801} (\bibinfo{year}{2007}{\natexlab{b}}), ISSN
  \bibinfo{issn}{0031-9007, 1079-7114},
  \urlprefix\url{http://link.aps.org/doi/10.1103/PhysRevLett.99.160801}.

\bibitem[{\citenamefont{Mow-Lowry et~al.}({2008})\citenamefont{Mow-Lowry,
  Mullavey, Gossler, Gray, and McClelland}}]{CoolingSpring2008}
\bibinfo{author}{\bibfnamefont{C.~M.} \bibnamefont{Mow-Lowry}},
  \bibinfo{author}{\bibfnamefont{A.~J.} \bibnamefont{Mullavey}},
  \bibinfo{author}{\bibfnamefont{S.}~\bibnamefont{Gossler}},
  \bibinfo{author}{\bibfnamefont{M.~B.} \bibnamefont{Gray}}, \bibnamefont{and}
  \bibinfo{author}{\bibfnamefont{D.~E.} \bibnamefont{McClelland}},
  \bibinfo{journal}{{Physical Review Letters}}
  \textbf{\bibinfo{volume}{{100}}}, \bibinfo{pages}{010801}
  (\bibinfo{year}{{2008}}), ISSN \bibinfo{issn}{{0031-9007}}.

\bibitem[{\citenamefont{Mahboob and Yamaguchi}(2008)}]{mahboob_bit_2008}
\bibinfo{author}{\bibfnamefont{I.}~\bibnamefont{Mahboob}} \bibnamefont{and}
  \bibinfo{author}{\bibfnamefont{H.}~\bibnamefont{Yamaguchi}},
  \bibinfo{journal}{Nature Nanotechnology} \textbf{\bibinfo{volume}{3}},
  \bibinfo{pages}{275} (\bibinfo{year}{2008}), ISSN \bibinfo{issn}{1748-3387,
  1748-3395},
  \urlprefix\url{http://www.nature.com/doifinder/10.1038/nnano.2008.84}.

\bibitem[{\citenamefont{Badzey et~al.}(2004)\citenamefont{Badzey,
  Zolfagharkhani, Gaidarzhy, and Mohanty}}]{badzey_controllable_2004}
\bibinfo{author}{\bibfnamefont{R.~L.} \bibnamefont{Badzey}},
  \bibinfo{author}{\bibfnamefont{G.}~\bibnamefont{Zolfagharkhani}},
  \bibinfo{author}{\bibfnamefont{A.}~\bibnamefont{Gaidarzhy}},
  \bibnamefont{and} \bibinfo{author}{\bibfnamefont{P.}~\bibnamefont{Mohanty}},
  \bibinfo{journal}{Applied Physics Letters} \textbf{\bibinfo{volume}{85}},
  \bibinfo{pages}{3587} (\bibinfo{year}{2004}), ISSN \bibinfo{issn}{00036951},
  \urlprefix\url{http://scitation.aip.org/content/aip/journal/apl/85/16/10.1063/1.1808507}.

\bibitem[{\citenamefont{Bagheri et~al.}(2011)\citenamefont{Bagheri, Poot, Li,
  Pernice, and Tang}}]{bagheri_dynamic_2011}
\bibinfo{author}{\bibfnamefont{M.}~\bibnamefont{Bagheri}},
  \bibinfo{author}{\bibfnamefont{M.}~\bibnamefont{Poot}},
  \bibinfo{author}{\bibfnamefont{M.}~\bibnamefont{Li}},
  \bibinfo{author}{\bibfnamefont{W.~P.~H.} \bibnamefont{Pernice}},
  \bibnamefont{and} \bibinfo{author}{\bibfnamefont{H.~X.} \bibnamefont{Tang}},
  \bibinfo{journal}{Nature Nanotechnology} \textbf{\bibinfo{volume}{6}},
  \bibinfo{pages}{726} (\bibinfo{year}{2011}), ISSN \bibinfo{issn}{1748-3387,
  1748-3395},
  \urlprefix\url{http://www.nature.com/doifinder/10.1038/nnano.2011.180}.

\bibitem[{\citenamefont{Gozzini et~al.}(1985)\citenamefont{Gozzini, Longo,
  Barbarino, Maccarrone, and Mango}}]{gozzini_light-pressure_1985}
\bibinfo{author}{\bibfnamefont{A.}~\bibnamefont{Gozzini}},
  \bibinfo{author}{\bibfnamefont{I.}~\bibnamefont{Longo}},
  \bibinfo{author}{\bibfnamefont{S.}~\bibnamefont{Barbarino}},
  \bibinfo{author}{\bibfnamefont{F.}~\bibnamefont{Maccarrone}},
  \bibnamefont{and} \bibinfo{author}{\bibfnamefont{F.}~\bibnamefont{Mango}},
  \bibinfo{journal}{JOSA B} \textbf{\bibinfo{volume}{2}}, \bibinfo{pages}{1841}
  (\bibinfo{year}{1985}),
  \urlprefix\url{http://www.osapublishing.org/abstract.cfm?uri=josab-2-11-1841}.

\bibitem[{\citenamefont{Mueller et~al.}({2008})\citenamefont{Mueller, Heugel,
  and Wang}}]{Mueller2008}
\bibinfo{author}{\bibfnamefont{F.}~\bibnamefont{Mueller}},
  \bibinfo{author}{\bibfnamefont{S.}~\bibnamefont{Heugel}}, \bibnamefont{and}
  \bibinfo{author}{\bibfnamefont{L.~J.} \bibnamefont{Wang}},
  \bibinfo{journal}{{Physical Review A}} \textbf{\bibinfo{volume}{{77}}},
  \bibinfo{pages}{031802(R)} (\bibinfo{year}{{2008}}), ISSN
  \bibinfo{issn}{{1050-2947}}.

\bibitem[{\citenamefont{Metzger et~al.}(2008)\citenamefont{Metzger, Ludwig,
  Neuenhahn, Ortlieb, Favero, Karrai, and
  Marquardt}}]{metzger_self-induced_2008}
\bibinfo{author}{\bibfnamefont{C.}~\bibnamefont{Metzger}},
  \bibinfo{author}{\bibfnamefont{M.}~\bibnamefont{Ludwig}},
  \bibinfo{author}{\bibfnamefont{C.}~\bibnamefont{Neuenhahn}},
  \bibinfo{author}{\bibfnamefont{A.}~\bibnamefont{Ortlieb}},
  \bibinfo{author}{\bibfnamefont{I.}~\bibnamefont{Favero}},
  \bibinfo{author}{\bibfnamefont{K.}~\bibnamefont{Karrai}}, \bibnamefont{and}
  \bibinfo{author}{\bibfnamefont{F.}~\bibnamefont{Marquardt}},
  \bibinfo{journal}{Physical Review Letters} \textbf{\bibinfo{volume}{101}},
  \bibinfo{pages}{133903} (\bibinfo{year}{2008}), ISSN
  \bibinfo{issn}{0031-9007, 1079-7114},
  \urlprefix\url{http://link.aps.org/doi/10.1103/PhysRevLett.101.133903}.

\bibitem[{\citenamefont{Hao et~al.}({2015})\citenamefont{Hao, JiangFang, Yong,
  and GengYu}}]{Hao_photothermal_bistable}
\bibinfo{author}{\bibfnamefont{F.}~\bibnamefont{Hao}},
  \bibinfo{author}{\bibfnamefont{D.}~\bibnamefont{JiangFang}},
  \bibinfo{author}{\bibfnamefont{L.}~\bibnamefont{Yong}}, \bibnamefont{and}
  \bibinfo{author}{\bibfnamefont{C.}~\bibnamefont{GengYu}},
  \bibinfo{journal}{{Science China-Physics Mechanics \& Astronomy}}
  \textbf{\bibinfo{volume}{{58}}} (\bibinfo{year}{{2015}}), ISSN
  \bibinfo{issn}{{1674-7348}}.

\bibitem[{\citenamefont{Buchmann et~al.}(2012)\citenamefont{Buchmann, Zhang,
  Chiruvelli, and Meystre}}]{buchmann_macroscopic_2012}
\bibinfo{author}{\bibfnamefont{L.~F.} \bibnamefont{Buchmann}},
  \bibinfo{author}{\bibfnamefont{L.}~\bibnamefont{Zhang}},
  \bibinfo{author}{\bibfnamefont{A.}~\bibnamefont{Chiruvelli}},
  \bibnamefont{and} \bibinfo{author}{\bibfnamefont{P.}~\bibnamefont{Meystre}},
  \bibinfo{journal}{Physical Review Letters} \textbf{\bibinfo{volume}{108}},
  \bibinfo{pages}{210403} (\bibinfo{year}{2012}), ISSN
  \bibinfo{issn}{0031-9007, 1079-7114},
  \urlprefix\url{http://link.aps.org/doi/10.1103/PhysRevLett.108.210403}.

\bibitem[{\citenamefont{Thompson et~al.}(2008)\citenamefont{Thompson, Zwickl,
  Jayich, Marquardt, Girvin, and Harris}}]{thompson_strong_2008}
\bibinfo{author}{\bibfnamefont{J.~D.} \bibnamefont{Thompson}},
  \bibinfo{author}{\bibfnamefont{B.~M.} \bibnamefont{Zwickl}},
  \bibinfo{author}{\bibfnamefont{A.~M.} \bibnamefont{Jayich}},
  \bibinfo{author}{\bibfnamefont{F.}~\bibnamefont{Marquardt}},
  \bibinfo{author}{\bibfnamefont{S.~M.} \bibnamefont{Girvin}},
  \bibnamefont{and} \bibinfo{author}{\bibfnamefont{J.~G.~E.}
  \bibnamefont{Harris}}, \bibinfo{journal}{Nature}
  \textbf{\bibinfo{volume}{452}}, \bibinfo{pages}{72} (\bibinfo{year}{2008}),
  ISSN \bibinfo{issn}{0028-0836, 1476-4687},
  \urlprefix\url{http://www.nature.com/doifinder/10.1038/nature06715}.

\bibitem[{\citenamefont{Jayich et~al.}(2008)\citenamefont{Jayich, Sankey,
  Zwickl, Yang, Thompson, Girvin, Clerk, Marquardt, and
  Harris}}]{jayich_dispersive_2008}
\bibinfo{author}{\bibfnamefont{A.~M.} \bibnamefont{Jayich}},
  \bibinfo{author}{\bibfnamefont{J.~C.} \bibnamefont{Sankey}},
  \bibinfo{author}{\bibfnamefont{B.~M.} \bibnamefont{Zwickl}},
  \bibinfo{author}{\bibfnamefont{C.}~\bibnamefont{Yang}},
  \bibinfo{author}{\bibfnamefont{J.~D.} \bibnamefont{Thompson}},
  \bibinfo{author}{\bibfnamefont{S.~M.} \bibnamefont{Girvin}},
  \bibinfo{author}{\bibfnamefont{A.~A.} \bibnamefont{Clerk}},
  \bibinfo{author}{\bibfnamefont{F.}~\bibnamefont{Marquardt}},
  \bibnamefont{and} \bibinfo{author}{\bibfnamefont{J.~G.~E.}
  \bibnamefont{Harris}}, \bibinfo{journal}{New Journal of Physics}
  \textbf{\bibinfo{volume}{10}}, \bibinfo{pages}{095008}
  (\bibinfo{year}{2008}), ISSN \bibinfo{issn}{1367-2630},
  \urlprefix\url{http://stacks.iop.org/1367-2630/10/i=9/a=095008?key=crossref.252fc7193dae0f986a14ff7316a8a9fe}.

\bibitem[{\citenamefont{Sankey et~al.}(2010)\citenamefont{Sankey, Yang, Zwickl,
  Jayich, and Harris}}]{sankey_strong_2010}
\bibinfo{author}{\bibfnamefont{J.~C.} \bibnamefont{Sankey}},
  \bibinfo{author}{\bibfnamefont{C.}~\bibnamefont{Yang}},
  \bibinfo{author}{\bibfnamefont{B.~M.} \bibnamefont{Zwickl}},
  \bibinfo{author}{\bibfnamefont{A.~M.} \bibnamefont{Jayich}},
  \bibnamefont{and} \bibinfo{author}{\bibfnamefont{J.~G.~E.}
  \bibnamefont{Harris}}, \bibinfo{journal}{Nature Physics}
  \textbf{\bibinfo{volume}{6}}, \bibinfo{pages}{707} (\bibinfo{year}{2010}),
  ISSN \bibinfo{issn}{1745-2473, 1745-2481},
  \urlprefix\url{http://www.nature.com/doifinder/10.1038/nphys1707}.

\bibitem[{\citenamefont{Mumford et~al.}(2015)\citenamefont{Mumford, O'Dell, and
  Larson}}]{mumford_dicke-type_2015}
\bibinfo{author}{\bibfnamefont{J.}~\bibnamefont{Mumford}},
  \bibinfo{author}{\bibfnamefont{D.~H.~J.} \bibnamefont{O'Dell}},
  \bibnamefont{and} \bibinfo{author}{\bibfnamefont{J.}~\bibnamefont{Larson}},
  \bibinfo{journal}{Annalen der Physik} \textbf{\bibinfo{volume}{527}},
  \bibinfo{pages}{115} (\bibinfo{year}{2015}), ISSN \bibinfo{issn}{00033804},
  \urlprefix\url{http://doi.wiley.com/10.1002/andp.201400105}.

\bibitem[{\citenamefont{Chan et~al.}(2011)\citenamefont{Chan, Alegre,
  Safavi-Naeini, Hill, Krause, Gröblacher, Aspelmeyer, and
  Painter}}]{chan_laser_2011}
\bibinfo{author}{\bibfnamefont{J.}~\bibnamefont{Chan}},
  \bibinfo{author}{\bibfnamefont{T.~P.~M.} \bibnamefont{Alegre}},
  \bibinfo{author}{\bibfnamefont{A.~H.} \bibnamefont{Safavi-Naeini}},
  \bibinfo{author}{\bibfnamefont{J.~T.} \bibnamefont{Hill}},
  \bibinfo{author}{\bibfnamefont{A.}~\bibnamefont{Krause}},
  \bibinfo{author}{\bibfnamefont{S.}~\bibnamefont{Gröblacher}},
  \bibinfo{author}{\bibfnamefont{M.}~\bibnamefont{Aspelmeyer}},
  \bibnamefont{and} \bibinfo{author}{\bibfnamefont{O.}~\bibnamefont{Painter}},
  \bibinfo{journal}{Nature} \textbf{\bibinfo{volume}{478}}, \bibinfo{pages}{89}
  (\bibinfo{year}{2011}), ISSN \bibinfo{issn}{0028-0836, 1476-4687},
  \urlprefix\url{http://www.nature.com/doifinder/10.1038/nature10461}.

\bibitem[{\citenamefont{Shimshoni et~al.}(2011)\citenamefont{Shimshoni, Morigi,
  and Fishman}}]{Shimshoni2011}
\bibinfo{author}{\bibfnamefont{E.}~\bibnamefont{Shimshoni}},
  \bibinfo{author}{\bibfnamefont{G.}~\bibnamefont{Morigi}}, \bibnamefont{and}
  \bibinfo{author}{\bibfnamefont{S.}~\bibnamefont{Fishman}},
  \bibinfo{journal}{Physical Review Letters} \textbf{\bibinfo{volume}{106}},
  \bibinfo{pages}{1} (\bibinfo{year}{2011}), ISSN \bibinfo{issn}{00319007}.

\end{thebibliography}

\end{document}